\documentstyle[prd,aps,epsf]{revtex}

\def\be{\begin{equation}}
\def\ee{\end{equation}}
\def\bea{\begin{eqnarray}}
\def\eea{\end{eqnarray}}
\def\simgt{\,\rlap{\lower 3.5 pt\hbox{$\mathchar \sim$}}\raise 1pt \hbox {$>$}\,}
\def\simlt{\,\rlap{\lower 3.5 pt\hbox{$\mathchar \sim$}}\raise 1pt \hbox {$<$}\,}
\begin{document}

\draft
\preprint{SWAT-367, UTCCP-P-130, UTHEP-464}

\tightenlines

\title{
\vspace*{-35pt}
{\normalsize \hfill {\sf SWAT-367, UTCCP-P-130, UTHEP-464}} \\
\vspace*{-6pt}
{\normalsize \hfill {\sf January \ 2003}} \\
Renormalization group improved action on anisotropic lattices
}

\author{S.~Ejiri$^{\rm a}$, K.~Kanaya$^{\rm b}$, Y.~Namekawa$^{\rm b}$,
 and T.~Umeda$^{\rm c}$}

\address{
$^{\rm a}$ Department of Physics, University of Wales Swansea,
    Singleton Park, Swansea, SA2 8PP, U.K., \\
$^{\rm b}$ Institute of Physics,
    University of Tsukuba, Tsukuba, Ibaraki 305-8571, Japan, \\
$^{\rm c}$ Center for Computational Physics,
    University of Tsukuba, Tsukuba, Ibaraki 305-8577, Japan
}

\date{\today}
\maketitle

\begin{abstract}
We study a block spin transformation in the SU(3) lattice gauge theory 
on anisotropic lattices to obtain Iwasaki's renormalization group 
improved action for anisotropic cases. 
For the class of actions with plaquette and $1\times2$ rectangular terms, 
we determine the improvement parameters as functions of the 
anisotropy $\xi= a_s/a_t$. 
We find that the program of improvement works well also on anisotropic 
lattices. 
From a study of an indicator which estimates the distance 
to the renormalized trajectory, 
we show that, for the range of the anisotropy $\xi \approx 1$--4, 
the coupling parameters previously determined for isotropic lattices 
improve the theory considerably.

\end{abstract}




\section{Introduction}
\label{sec:intro}

Improvement and anisotropy are two key ingredients of the recent 
developments in lattice QCD. 
In QCD with dynamical quarks, 
improvement of the lattice theory is essential to perform 
a continuum extrapolation of light hadron spectra 
within the computer power currently available \cite{cppacsFull,comparative}. 
At finite temperatures, 
the expected O(4) scaling around the two-flavor chiral transition point 
is reproduced on the lattice only with improved Wilson-type quarks 
\cite{tsukuba97,cppacs00}.
Improved actions are applied also for staggered-type quarks 
to reduce lattice artifacts \cite{milc,p4action}. 
However, 
in order to perform a continuum extrapolation of thermodynamic 
quantities, we need to further increase the temporal lattice size $N_t$
\cite{milc97,cppacs01}. 
This requires quite large spatial lattice sizes to keep the system 
close to the thermodynamic limit, and the task slightly exceeds the 
current limit of the computer power for QCD with dynamical quarks 
\cite{ejiri01}.

Recently, we proposed to apply anisotropic lattices for reducing the 
computational demand in thermal QCD \cite{name01}. 
Because the dominant part of the lattice artifacts in the equation
of state (EOS) is due to the finite temporal cutoff, an anisotropic lattice 
with a larger temporal cutoff will provide us with an efficient way to 
calculate thermodynamic quantities.
We tested the idea for the case of the SU(3) gauge theory with 
the standard one-plaquette action.
From a series of simulations at $N_t/\xi=4$, 5 and 6 with the anisotropy 
$\xi \equiv a_s / a_t = 2$, 
where $a_s$ and $a_t$ are the spatial and temporal lattice spacings, 
we find that the lattice artifacts in EOS are much smaller than 
those on the corresponding isotropic lattice, and the leading scaling 
relation is satisfied from the coarsest lattice. 
This enabled us to perform a well-controlled continuum extrapolation 
of EOS in QCD. 
Anisotropic lattices have been employed also 
to study transport coefficients and temporal correlation functions 
in finite temperature QCD \cite{sakai,taro,umeda}. 
In these studies, anisotropy was introduced to obtain more data points 
for temporal correlation functions.
At zero temperature, anisotropic lattices have been employed to study 
charmonium states \cite{klassen,bali,okamoto}, heavy hybrids \cite{manke}, 
glueballs \cite{morningstar}, and also the pion scattering length \cite{liu}. 

Combination of the ideas of improvement and anisotropy is not straightforward,
however. 
The main difficulty is the large number of coupling parameters in improved 
actions on anisotropic lattices. 
Even in the simplest case of the renormalization-group (RG) improved gauge 
action by Iwasaki \cite{iwasaki}, which contains plaquette and $1\times2$ 
rectangular terms only, we have 5 parameters on anisotropic lattices, 
instead of 2 for the isotropic case. 
We have to fix them as functions of two parameters, the gauge coupling $\beta$
which controls the overall scale, and the anisotropy $\xi$. 
Because the redundant parameters have no physical effects in the continuum 
limit, they have to be determined through a
requirement of improvement, i.e. minimizing lattice artifacts in physical 
observables away from the continuum limit.

Concrete form of the dependence on the scale and anisotropy in the coupling 
parameters is important for a calculation of thermodynamic quantities 
\cite{karsch81,burgers88,klassen98,ejiri98,engels00}. 
In a Symanzik-type improvement program, it is easy to see that, 
at the tree level of perturbation theory, 
the coupling parameters are independent of $\xi$. 
Accordingly, studies of finite temperature Symanzik-type improved actions 
on anisotropic lattices have been done assuming isotropic improvement 
parameters \cite{alford98,morningstar97,alford01}. 
When we improve the theory beyond the tree level, we have to take into account 
the $\xi$-dependences in the coupling parameters.
Isotropic parameters have been adopted also in a study of the RG-improved 
action on anisotropic lattices \cite{sakai}, 
however, without justifying the choice.

In this paper, we study the anisotropic improvement parameters for the 
RG-improved gauge action.
Following Iwasaki's program of RG-improvement using a block spin 
transformation, we determine the values of coupling parameters 
which minimize the lattice discretization errors.
After a brief explanation of the RG-improved action in Sec.~\ref{sec:RG}, 
the anisotropic gauge action we study is defined in 
Sec.~\ref{sec:action}. 
We then study Wilson loops under a block spin transformation
in Sec.~\ref{sec:wilson}. 
In Sec.~\ref{sec:ARG}, RG-improved actions on anisotropic lattices are 
determined and a practical choice of the improved action for numerical 
simulations is discussed.
We conclude in Sec.~\ref{sec:conc}.

\section{Program for RG-improved action}
\label{sec:RG}

Various lattice actions are expected to belong to a common 
universality class having the same continuum limit. 
For the SU(3) gauge theory, 
a lattice action may contain, for example, $1\times2$ rectangular loops, 
$2\times2$ squares, etc., in addition to the conventional plaquettes. 
One combination $\beta=2N_c/g^2$ of the coupling parameters is the relevant 
parameter which reflects the freedom of the lattice spacing, 
while other coupling parameters are redundant in the continuum limit. 
The objective of improvement is to find the values of redundant parameters 
for which physical observables from a coarse lattice 
are closest to their continuum values. 

\begin{figure}[tb]
  \begin{center}
    \leavevmode
    \epsfxsize=12cm \epsfbox{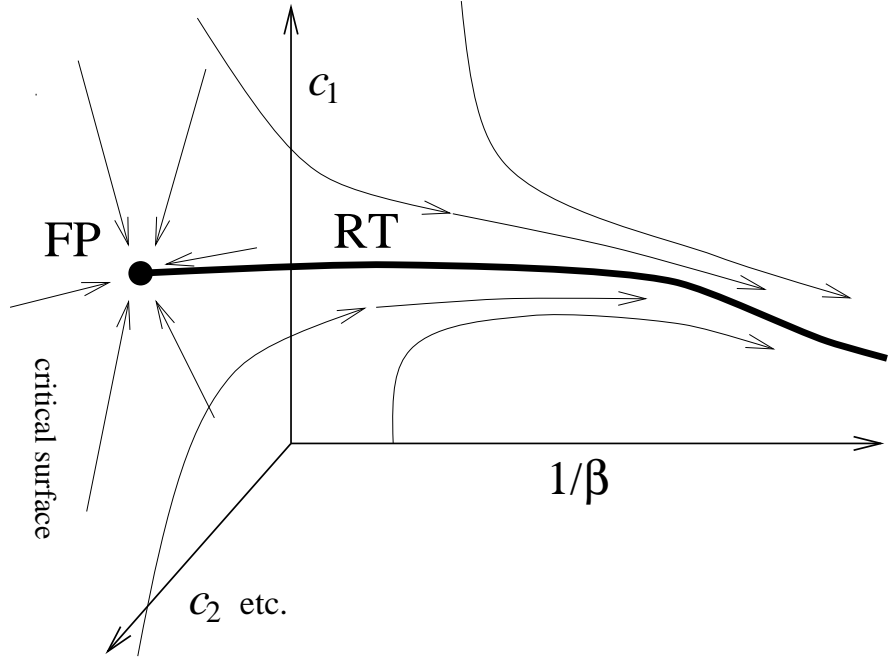}
    \caption{RG flow and the renormalized trajectory for the SU(3) 
gauge theory in infinite dimensional coupling parameter space.}
    \label{fig:RG}
  \end{center}
\end{figure}

In order to discuss the improvement of a lattice action, we consider 
RG flows of a block spin transformation as shown in Fig.~\ref{fig:RG}.
The block spin transformation halves the correlation length in lattice units 
but does not change the long-range properties of the system. 
Then, the coupling parameter moves toward smaller $\beta$ corresponding 
to the correlation length becoming shorter. 
In this figure, $c_1$, $c_2$, etc.\ denote the redundant coupling 
parameters, and the points on the $1/\beta$ axis correspond to 
the standard one plaquette actions.
The hyperplane at $\beta=\infty$ $(g=0)$, on which the correlation length diverges 
and the continuum limit can be taken, is called the critical surface. 
In this surface, 
the coupling parameter does not go out of the surface of $\beta=\infty$ 
under the block spin transformation, 
since the correlation length after the block spin 
transformation is also infinity, which means that an RG flow around 
$\beta=\infty$ runs parallel to the critical surface, except in the vicinity 
of a fixed point (FP), at which a coupling parameter does not change 
under the block spin transformation. 
Therefore, an RG flow can connect only at FP to the critical surface. 

Moreover, because RG flows can be regarded as lines of constant physics, 
the distance between each RG flow becomes wider as $\beta$ increases 
corresponding to physical quantities becoming more insensitive to 
the redundant coupling parameters, $c_1$, $c_2$, etc., 
as the continuum limit is approached, and only one RG flow which has 
properties in the continuum limit, called the renormalized trajectory (RT), 
can connect to the critical surface at FP 
by infinite block spin transformations.

The actions on the RT are ``perfect actions'' which reproduce 
continuum properties from the shortest distances on the lattice 
\cite{hasenfratz94}.
If an infinite number of coupling parameters are admitted, a perfect action
is a goal of improvement.
In reality, we are forced to keep the interactions as simple as possible 
in numerical simulations. Hence to find the nearest point to RT 
in the restricted coupling parameter space is the problem in practice. 
Iwasaki applied the program of improvement to the SU($N_c$) gauge theory 
in the weak coupling limit for the case of the action with plaquette 
and $1 \times 2$ rectangular terms \cite{iwasaki}.
He found the nearest point to the fixed point in the coupling parameter 
subspace by calculating Wilson loops perturbatively on 
the lattice consisting of blocked link 
valuables after block spin transformations.
Another approach of an RG-improved action 
is the classical perfect action approach in \cite{hasenfratz94}, 
which is suitable for increasing the coupling parameters.
See \cite{taro98} for a trial to include quantum corrections.
A classically perfect action on anisotropic lattices was studied in 
\cite{ruefenacht}.


\section{Gauge theory on anisotropic lattice}
\label{sec:action}

On isotropic lattices, the RG-improved gauge action by Iwasaki \cite{iwasaki},
which consists of plaquettes $W^{(1 \times 1)}$ and $1\times2$ rectangular
loops $W^{(1 \times 2)}$, is defined by
\be
 S_{imp} = -\beta \left\{ \sum_{x,\, \mu > \nu} 
      c_{0} W_{\mu\nu}^{(1 \times 1)}(x) 
    + \sum_{x,\, \mu \neq \nu} c_{1} W_{\mu\nu}^{(1 \times 2)}(x) \right\}, 
\label{eq:Iwasaki}
\ee
where $c_0 + 8 c_1 = 1$ for normalization and 
$c_1 = - 0.331$ ($-0.293$) to optimize the action after one (two) block 
spin transformation(s) (see below for details).
This action has been shown to lead to better rotation symmetry of heavy quark 
potential than the standard one plaquette action \cite{comparative,RGpot}, 
and to suppress lattice artifacts associated with Wilson-type quarks 
at finite temperature \cite{tsukuba97}. 
This action was also reported to be efficient in suppressing chiral violations 
in domain-wall quarks \cite{cppacsDW}. 
In two-flavor full QCD with clover-improved Wilson quarks, 
the first systematic studies of the light hadron spectrum 
\cite{cppacsFull} and 
the finite temperature equation of state with Wilson-type quarks 
\cite{cppacs00,cppacs01} have been carried out.

The generalization of Iwasaki's action to an anisotropic lattice is  
given by
\begin{eqnarray}
 S &=& -\beta_{s} \left\{ \sum_{x,\, i > j} 
      c_{0}^{s} W_{i j}^{(1 \times 1)}(x) 
    + \sum_{x,\, i \neq j} c_{1}^{s} W_{i j}^{(1 \times 2)}(x) \right\} 
\nonumber \\ 
   & & -\beta_{t} \left\{ \sum_{x,\, k} 
      c_{0}^{t} W_{k 4}^{(1 \times 1)}(x) 
    + \sum_{x,\, k} [ c_{1}^{t} W_{k 4}^{(2 \times 1)}(x)
                 + c_{2}^{t} W_{k 4}^{(1 \times 2)}(x) ] \right\}
\label{eq:gaction}
\end{eqnarray}
where we set $c_0^s+8c_1^s=1$ and $c_0^t+4c_1^t+4c_2^t=1$ for normalization.
This form of the tree-level Symanzik-improved action with mean field 
improvement was studied in \cite{alford98,morningstar97,alford01}. 
Here, we study the RG transformation of this action to obtain an RG-improved 
action.
\footnote{
Because our action contains couplings extending over two time slices, 
unphysical higher lying states may contaminate correlation functions at 
short distances comparable to the extent of the action,
as observed, e.g., in a study of glueball spectrum using a Symanzik 
improved gauge action \cite{morningstar96}.
Although these unphysical states do not affect physical properties at
long distances, a caution is required when we have to study
short distance correlators to extract physical quantities.
}

Let us denote the lattice spacing in the $\mu$-direction as $a_{\mu}$ and 
the lattice size as $N_{\mu}$. 
We consider the case $a_1=a_2=a_3 \equiv a_s$ and $a_4 \equiv a_t$, 
and $N_1=N_2=N_3 \equiv N_s$ and $N_4 \equiv N_t$ with 
sufficiently large $N_s$ and $N_t$.
Identifying the gauge field by $U_{\mu}(x)=\exp[iga_{\mu} A_{\mu}(x)]$, 
the conventional gauge action is recovered in the classical continuum limit
when
\begin{eqnarray}
\beta_{s}=\frac{2 N_c}{g^2 \xi}, \hspace{5mm} 
\beta_{t}=\frac{2 N_c \xi}{g^2},
\label{eq:betast}
\end{eqnarray}
where $\xi=a_s/a_t$.

We perform the Fourier transformation of $A_\mu(x)$ by 
\begin{eqnarray}
A_{\mu}(x)=\int_k {\rm e}^{ikx +ik_{\mu}a_{\mu}/2} \tilde{A}_{\mu}(k),
\label{eq:four}
\end{eqnarray}
where $\int_k \equiv \frac{1}{\sqrt{N_s^3 N_t}} \prod_{\mu=1}^{4} 
\sum_{k_{\mu}}$,
$k_{\mu}=\frac{2 \pi j_{\mu}}{N_{\mu} a_{\mu}}$, and 
$j_{\mu}$ is integer.
Then, the action reads
\begin{eqnarray}
 S &=& \frac{1}{2} a_s^3 a_t \prod_{\mu=1}^{4} \sum_{k_{\mu}} \left\{
  \sum_{i<j,a} \{1-c_1^s a_s^2 (\hat{k}_i^2 + \hat{k}_j^2) \} 
  \tilde{f}_{ij}^a (k) \tilde{f}_{ij}^a (-k) \right. + 
\nonumber \\
&&  \left. \sum_{i,a} (1-c_1^t a_s^2 \hat{k}_i^2 - c_2^t a_t^2 \hat{k}_4^2) 
  \tilde{f}_{i4}^a (k) \tilde{f}_{i4}^a (-k) \right\} + O[\tilde{A}^3],
\label{eq:s0a2}
\end{eqnarray}
with
$\hat{k}_{\mu}=(2/a_{\mu}) \sin (k_{\mu} a_{\mu}/2)$ and 
$\tilde{f}_{\mu \nu} (k) 
= i(\hat{k}_{\mu} \tilde{A}_{\nu}(k) - \hat{k}_{\nu} \tilde{A}_{\mu}(k))$.
We adopt the lattice Lorentz gauge by adding the gauge fixing term:
\begin{eqnarray}
S_{\rm gf} &=& 
a_s^3 a_t \sum_x {\rm tr} \left[\sum_{\mu} \Delta_{\mu} A_{\mu}(x)\right]^2, 
\\
\Delta_{\mu}f(x) &=& \{f(x) - U^{\dagger}_{\mu} 
f(x-a_{\mu} \hat{\mu}) U_{\mu} (x-a_{\mu} \hat{\mu})\}/a_{\mu}
\end{eqnarray}

In order to simplify the notation, we redefine lattice momenta and the 
gauge field absorbing the lattice spacings as
$k_{\mu} a_{\mu} \to k_{\mu}$, $\hat{k}_{\mu} a_{\mu} \to \hat{k}_{\mu}$, 
and $\tilde{A}_{\mu} a_{\mu} \to \tilde{A}_{\mu}$, in the following. 
Then, the lattice propagator, 
$\langle \tilde{A}_{\mu}^a (k) \tilde{A}_{\nu}^b (k') \rangle 
= \delta_{a,b} \delta (k+k') D_{\mu \nu}(k)$, is given by
\begin{eqnarray}
D_{ij}^{-1} (k) &=& \sum_{l=1}^{3} \frac{1}{\xi} q_{li} (k) \hat{k}_{l}^2 
\delta_{ij} + \xi q_{4i} (k) \hat{k}_4^2 \delta_{ij} - \frac{1}{\xi} 
(q_{ij} (k) -1) \hat{k}_i \hat{k}_j, \nonumber \\
D_{i4}^{-1} (k) &=& - \xi (q_{i4} (k) -1) \hat{k}_i \hat{k}_4, \hspace{5mm} 
D_{4i}^{-1} (k) = - \xi (q_{4i} (k) -1) \hat{k}_4 \hat{k}_i, \nonumber \\
D_{44}^{-1} (k) &=& \sum_{l=1}^{3} \xi q_{l4} (k) \hat{k}_{l}^2 
+ \xi^3 \hat{k}_4^2,
\label{eq:dinv}
\end{eqnarray}
where
\begin{eqnarray}
q_{ij} (k) &=& 1 - c_1^s (\hat{k}_i^2 + \hat{k}_j^2) \hspace{5mm} 
{\rm for} \hspace{2mm} i \neq j \,(=1,2,3), \nonumber \\
q_{i4} (k) &=& q_{4i}(k) = 1 - c_1^t \hat{k}_i^2 - c_2^t \hat{k}_4^2
\end{eqnarray}
with $q_{\mu \mu}(k)= 0$.

We consider the following Wilson loops,
\begin{eqnarray}
W_{\mu \nu} (1 \times 1) &=& (1/N_c) {\rm tr}[U_{\mu}(x) U_{\nu}(x+\hat{\mu}) 
U^{\dagger}_{\mu}(x+\hat{\nu}) U^{\dagger}_{\nu}(x)], 
\label{eq:plaq} \\
W_{\mu \nu} (2 \times 1) &=& (1/N_c) {\rm tr}[U_{\mu}(x) U_{\mu}(x+\hat{\mu}) 
U_{\nu}(x+2\hat{\mu}) U^{\dagger}_{\mu}(x+\hat{\mu}+\hat{\nu}) 
U^{\dagger}_{\mu}(x+\hat{\nu}) U^{\dagger}_{\nu}(x)], 
\label{eq:rect} \\
W_{\mu \nu \rho} ({\rm chair}) &=& (1/N_c) {\rm tr}[U_{\mu}(x) 
U_{\nu}(x+\hat{\mu}) 
U_{\rho}(x+\hat{\mu}+\hat{\nu}) U^{\dagger}_{\mu}(x+\hat{\nu}+\hat{\rho}) 
U^{\dagger}_{\rho}(x+\hat{\nu}) U^{\dagger}_{\nu}(x)], 
\label{eq:chair} \\
W_{\mu \nu \rho} ({\rm 3~dim}) &=& (1/N_c) {\rm tr}[U_{\mu}(x) 
U_{\nu}(x+\hat{\mu}) 
U_{\rho}(x+\hat{\mu}+\hat{\nu}) U^{\dagger}_{\mu}(x+\hat{\nu}+\hat{\rho}) 
U^{\dagger}_{\nu}(x+\hat{\rho}) U^{\dagger}_{\rho}(x)].
\label{eq:3dim}
\end{eqnarray}
To the leading order of perturbation theory, we get \cite{iwasaki,weisz83}
\begin{eqnarray}
\langle W(C) \rangle &\equiv & 1-g^2 \frac{N_c^2 -1}{4N_c} F(C),
\\
F_{\mu \nu} (I \times J) &=& 
(N_s^3 N_t)^{-1} \prod_{\rho=1}^{4} \sum_{k_{\rho}} 
\left( \frac{\sin (I k_{\mu}/2)}{\sin (k_{\mu}/2)}
\frac{\sin (J k_{\nu}/2)}{\sin (k_{\nu}/2)} \right)^2 
D_{\mu \nu, \mu \nu} (k)
\\
F_{\mu \nu \rho} ({\rm chair}) &=& 
(N_s^3 N_t)^{-1} \prod_{\sigma=1}^{4} \sum_{k_{\sigma}} 
[D_{\mu \nu, \mu \nu} (k) + D_{\mu \rho, \mu \rho} (k) 
\nonumber \\
&&
- \frac{1}{2} (\hat{k}_{\mu}^2 \hat{k}_{\nu} \hat{k}_{\rho} D_{\nu \rho}(k)
+ \hat{k}_{\nu}^2 \hat{k}_{\rho}^2 D_{\mu \mu} (k)
- \hat{k}_{\mu} \hat{k}_{\nu}^2 \hat{k}_{\rho} D_{\mu \rho} (k)
- \hat{k}_{\mu} \hat{k}_{\nu} \hat{k}_{\rho}^2 D_{\mu \nu} (k))], 
\\
F_{\mu \nu \rho} ({\rm 3~dim}) &=& 
(N_s^3 N_t)^{-1} \prod_{\sigma=1}^{4} \sum_{k_{\sigma}} 
[(1 - \hat{k}_{\rho}^2 /4)D_{\mu \nu, \mu \nu} (k) 
+ (1 - \hat{k}_{\mu}^2 /4)D_{\nu \rho, \nu \rho} (k) 
+ (1 - \hat{k}_{\nu}^2 /4)D_{\rho \mu, \rho \mu} (k) ],
\end{eqnarray}
for $SU(N_c)$ gauge theory, where
$D_{\mu \nu, \mu \nu}(k) = \hat{k}_{\mu}^2 D_{\nu \nu} (k) 
- \hat{k}_{\mu} \hat{k}_{\nu} D_{\nu \mu} (k)
- \hat{k}_{\nu} \hat{k}_{\mu} D_{\mu \nu} (k)
+ \hat{k}_{\nu}^2 D_{\mu \mu} (k) $.

\section{Block spin transformation}
\label{sec:wilson}

The purpose of this study is to find a fixed point, 
at which the parameters in the action do not change. 
As seen in the previous section, $F(C)$ are functions of the redundant 
parameters $c_i^{s/t}$, hence, if a block spin transformation is performed 
in the vicinity of the fixed point, the values of $F(C)$ should not change. 
In this section, we calculate Wilson loops on the blocked lattice 
after block spin transformations in the $g \to 0$ limit 
and discuss the fixed point.

Following Iwasaki \cite{iwasaki}, we consider a simple block spin 
transformation from $N_{\rm BS}$-th to $(N_{\rm BS}+1)$-th blocking, 
of the form 
\begin{eqnarray}
A_{\mu}^{(N_{\rm BS}+1)} (n') 
= \frac{1}{8} \sum_{n \in n'} A_{\mu}^{(N_{\rm BS})} (n),
\label{eq:block}
\end{eqnarray}
where we block $2^4$ links at the sites 
$n = 2n' + \sum_{\mu} \epsilon_{\mu} \hat{\mu}$, $(\epsilon_\mu=0,1)$ 
to 1 link at $n'$ on the blocked lattice.
The lattice spacings change from $a_{\mu}$ to $2a_{\mu}$ by this 
transformation, while the anisotropy remains the same.
Note that the scale factor 2 is multiplied on the right hand side of 
Eq.(\ref{eq:block}) to scale back to the original lattice spacings, 
so that the relevant coupling $g$ remains constant.

Link variables on the blocked lattice are defined by 
$U_{\mu}^{(N_{\rm BS})} (n) = \exp ( iga_{\mu} A_{\mu}^{(N_{\rm BS})} (n) )$. 
Wilson loops consisting of blocked links are given by
\cite{iwasaki} 
\begin{eqnarray}
\langle W^{(N_{\rm BS})} (C) \rangle 
&\equiv & 1-g^2 \frac{N_c^2 -1}{4N_c} F^{(N_{\rm BS})} (C), 
\\
F_{\mu \nu}^{(N_{\rm BS})} (I \times J) &=& 
(N_s^3 N_t)^{-1} \prod_{\rho=1}^{4} \sum_{k_{\rho}} \left(
\frac{\sin (I k_{\mu}^{(N_{\rm BS})}/2)}{\sin (k_{\mu}^{(N_{\rm BS})}/2)}
\frac{\sin (J k_{\nu}^{(N_{\rm BS})}/2)}{\sin (k_{\nu}^{(N_{\rm BS})}/2)} 
\right)^2 
D_{\mu \nu, \mu \nu}^{(N_{\rm BS})}(k) H^{(N_{\rm BS})} (k), 
\label{eq:fbij} \\
F_{\mu \nu \rho}^{(N_{\rm BS})} ({\rm chair}) &=& 
(N_s^3 N_t)^{-1} \prod_{\sigma=1}^{4} 
\sum_{k_{\sigma}} \left[D_{\mu \nu, \mu \nu}^{(N_{\rm BS})} (k) 
+ D_{\mu \rho, \mu \rho}^{(N_{\rm BS})} (k) 
- \frac{1}{2} \left( (\hat{k}_{\mu}^{(N_{\rm BS})})^2 
\hat{k}_{\nu}^{(N_{\rm BS})} 
\hat{k}_{\rho}^{(N_{\rm BS})} D_{\nu \rho}^{(N_{\rm BS})} (k) \right. \right.
\nonumber \\ 
&& + (\hat{k}_{\nu}^{(N_{\rm BS})} \hat{k}_{\rho}^{(N_{\rm BS})})^2 
D_{\mu \mu}^{(N_{\rm BS})} (k) 
- \hat{k}_{\mu}^{(N_{\rm BS})} (\hat{k}_{\nu}^{(N_{\rm BS})})^2 
\hat{k}_{\rho}^{(N_{\rm BS})} D_{\mu \rho}^{(N_{\rm BS})} (k)
\nonumber \\
&& \left. \left. - \hat{k}_{\mu}^{(N_{\rm BS})} \hat{k}_{\nu}^{(N_{\rm BS})} 
(\hat{k}_{\rho}^{(N_{\rm BS})})^2 
D_{\mu \nu}^{(N_{\rm BS})} (k)) \right) \right] H^{(N_{\rm BS})} (k), 
\label{eq:fbch} \\
F_{\mu \nu \rho}^{(N_{\rm BS})} ({\rm 3~dim}) &=& 
(N_s^3 N_t)^{-1} \prod_{\sigma=1}^{4} 
\sum_{k_{\sigma}} \left[ 
\left(1 - \frac{(\hat{k}_{\rho}^{(N_{\rm BS})})^2}{4} \right) 
D_{\mu \nu, \mu \nu}^{(N_{\rm BS})} (k) 
\right. \nonumber \\
&& \left.
+ \left(1 - \frac{(\hat{k}_{\mu}^{(N_{\rm BS})})^2}{4} \right) 
D_{\nu \rho, \nu \rho}^{(N_{\rm BS})} (k) 
+ \left(1 - \frac{(\hat{k}_{\nu}^{(N_{\rm BS})})^2}{4} \right) 
D_{\rho \mu, \rho \mu}^{(N_{\rm BS})} (k) 
\right] H^{(N_{\rm BS})} (k), 
\label{eq:fb3d}
\end{eqnarray}
to the leading order, where
\begin{eqnarray}
k_{\mu}^{(N_{\rm BS})} &=& 2^{N_{\rm BS}} k_{\mu}, \hspace{5mm} 
\hat{k}_{\mu}^{(N_{\rm BS})} = 2 \sin (k_{\mu}^{(N_{\rm BS})} /2), 
\\
H^{(N_{\rm BS})} (k) &=& 
\prod_{M=0}^{N_{\rm BS}-1} \frac{1}{4} \prod_{\mu=1}^{4} 
(1+ \cos (2^M k_{\mu})) 
\nonumber \\
D_{\mu \nu, \mu \nu}^{(N_{\rm BS})} (k) &=& 
(\hat{k}_{\mu}^{(N_{\rm BS})})^2 
D_{\nu \nu} (k) 
+ (\hat{k}_{\nu}^{(N_{\rm BS})})^2 D_{\mu \mu} (k) \nonumber 
\\ 
&&
- 2 \hat{k}_{\mu}^{(N_{\rm BS})} \hat{k}_{\nu}^{(N_{\rm BS})} 
\cos((2^{N_{\rm BS}-1} - 1/2) k_{\mu})
 \cos((2^{N_{\rm BS}-1} - 1/2) k_{\nu}) D_{\mu \nu} (k) .
\label{eq:hbs}
\end{eqnarray}
The derivation of Eqs.(\ref{eq:fbij}), (\ref{eq:fbch}) and (\ref{eq:fb3d}) 
is given in Appendix A.

In Table~\ref{tab:wl}, we list the numerical results of $F^{(N_{\rm BS})}$ 
for the case of the standard one plaquette action.
We find that the values of $F^{(N_{\rm BS})}$ approach to specific 
values in the $N_{\rm BS} \to \infty$ limit. 

Wilson loops in the limit of infinite $N_{\rm BS}$ can be evaluated as follows.
At long distances, the gauge propagator should behave like 
\begin{eqnarray}
\langle A_{\mu}^a (x) A_{\nu}^b (0) \rangle =
\frac{1}{4 \pi^2} \frac{\delta_{\mu \nu} \delta_{a, b}}
{x_1^2 + x_2^2 + x_3^2 + x_4^2} + O(1/x^4)
\end{eqnarray}
in physical unit.
In lattice units, it reads
\begin{eqnarray}
\label{eq:infpro}
\langle A_{\mu}^a (n) A_{\nu}^b (0) \rangle 
&=& \delta_{\mu \nu} \delta_{a, b}
f_{\mu} (n) + O(1/n^4), 
\\
&& \hspace{-25mm} f_{i} (n) = \frac{1}{4 \pi^2} 
\frac{1}{n_1^2 + n_2^2 + n_3^2 + \xi^{-2} n_4^2}, \hspace{5mm} 
f_{4} (n) = \frac{1}{4 \pi^2} 
\frac{1}{\xi^2 n_1^2 + \xi^2 n_2^2 + \xi^2 n_3^2 + n_4^2}. 
\nonumber
\end{eqnarray}
The non-leading term of the right hand side of Eq.(\ref{eq:infpro}) 
do not contribute to the expectation value in the 
$N_{\rm BS} \to \infty$ limit \cite{iwasaki}.
Hence we can neglect the higher order terms. 
Then the resulting Wilson loops do not depend on the improvement 
parameters $c_i^{s/t}$ in the original action, 
since the leading term does not depend on them.

Now, the $(I \times J)$ rectangular Wilson loops in the limit 
$N_{\rm BS} \to \infty$ are given by 
\begin{eqnarray}
F_{\mu \nu}^{(\infty)} (I \times J) &=& 
\lim_{N_{\rm BS} \to \infty} \left( 
\frac{1}{8^{N_{\rm BS}}} \right)^2 \sum_{m,n} 
\left[ 2I f_{\mu} (m-n) + 2J f_{\nu} (m-n) 
+ 4 \sum_{k=1}^{I-1} (I-k) f_{\mu} (2^{N_{\rm BS}} k \hat{\mu} +m-n) 
 \right. \nonumber \\ && 
+ 4 \sum_{k=1}^{J-1} (J-k) f_{\nu} (2^{N_{\rm BS}} k \hat{\nu} +m-n) 
- 2I f_{\mu} (2^{N_{\rm BS}} J \hat{\nu} +m-n) 
- 2J f_{\nu} (2^{N_{\rm BS}} I \hat{\mu} +m-n) 
\nonumber \\ 
&& \left.
- 4 \sum_{k=1}^{I-1} (I-k) f_{\mu} (2^{N_{\rm BS}} (k \hat{\mu} 
+ J \hat{\nu}) +m-n) 
- 4 \sum_{k=1}^{J-1} (J-k) f_{\nu} (2^{N_{\rm BS}} (k \hat{\nu} 
+ I \hat{\mu}) +m-n) \right]
\\ 
&=& 2 \prod_{\rho =1}^{4} \int_0^1 {\rm d}x_{\rho} (1 - x_{\rho})
\left[I \tilde{f}_{\mu} (0) + J \tilde{f}_{\nu} (0) 
+ 2 \sum_{k=1}^{I-1} (I-k) \tilde{f}_{\mu} (k \hat{\mu}) 
+ 2 \sum_{k=1}^{J-1} (J-k) \tilde{f}_{\nu} (k \hat{\nu}) \right.
\nonumber \\ 
&& \left.
- I \tilde{f}_{\mu} (J \hat{\nu}) - J \tilde{f}_{\nu} (I \hat{\mu})
- 2 \sum_{k=1}^{I-1} (I-k) \tilde{f}_{\mu} (k \hat{\mu} + J \hat{\nu}) 
- 2 \sum_{k=1}^{J-1} (J-k) \tilde{f}_{\nu} (k \hat{\nu} + I \hat{\mu}) 
\right],
\end{eqnarray}
where
\begin{eqnarray}
\tilde{f}_{i=1,2,3} (n) &=& \frac{1}{4 \pi^2} \prod_{\mu=1}^{4} 
\sum_{\epsilon_{\mu} = \{-1,1 \}}
\frac{1}{(n_1-\epsilon_1x_1)^2 +(n_2-\epsilon_2x_2)^2 
+(n_3-\epsilon_3x_3)^2 + \xi^{-2} (n_4-\epsilon_4x_4)^2}, \\
\tilde{f}_{4} (n) &=& \frac{1}{4 \pi^2} \prod_{\mu=1}^{4} 
\sum_{\epsilon_{\mu} = \{-1,1 \}}
\frac{1}{\xi^2 [ (n_1-\epsilon_1x_1)^2 + (n_2-\epsilon_2x_2)^2 
+ (n_3-\epsilon_3x_3)^2] + (n_4-\epsilon_4x_4)^2}.
\end{eqnarray}
Here, we have used 
$\lim_{N \to \infty} \sum_{n=1}^{2^N} f(n/2^N) = \int_0^1 {\rm d}x f(x)$
and a relation
$\int_0^1 {\rm d}x \int_0^1 {\rm d}y f(x-y) 
= \int_0^1 {\rm d}x (1-x) [f(x) + f(-x)] $. 
Similarly, we obtain,
\begin{eqnarray}
F^{(\infty)}_{\mu \nu \rho} ({\rm chair}) 
&=& 2 \prod_{\sigma =1}^{4} \int_0^1 {\rm d}x_{\sigma} (1 - x_{\sigma})
(\tilde{f}_{\mu} (0) + \tilde{f}_{\nu} (0) + \tilde{f}_{\rho} (0) 
- \tilde{f}_{\nu} (\hat{\mu}) - \tilde{f}_{\rho} (\hat{\mu})  
- \tilde{f}_{\mu} (\hat{\nu}+\hat{\rho})), \\
F^{(\infty)}_{\mu \nu \rho} ({\rm 3~dim}) 
&=& 2 \prod_{\sigma =1}^{4} \int_0^1 {\rm d}x_{\sigma} (1 - x_{\sigma})
(\tilde{f}_{\mu} (0) + \tilde{f}_{\nu} (0) + \tilde{f}_{\rho} (0) 
- \tilde{f}_{\mu} (\hat{\nu}+\hat{\rho}) 
- \tilde{f}_{\nu} (\hat{\rho}+\hat{\mu}) 
- \tilde{f}_{\rho} (\hat{\mu}+\hat{\nu})). 
\end{eqnarray}
We also denote $F^{(\infty)}$ in Table~\ref{tab:wl}.

From the behavior that $F(C)^{\rm (N_{BS})}$ converges monotonically 
to $F(C)^{(\infty)}$ which is independent of the coupling parameters 
$c_i^{s/t}$ in the original action, i.e. the starting point of RG flow, 
if a block spin transformation is performed from the point 
at which $F(C)^{\rm (N_{BS})}$ is already $F(C)^{(\infty)}$, 
we expect that the value of $F(C)^{\rm (N_{BS})}$ does not change anymore, 
which is the property at a fixed point. 
Therefore we can identify the fixed point by how close the value of 
$F(C)^{\rm (N_{BS})}$ is to $F(C)^{(\infty)}$. 

Notice that the property of $F(C)^{\rm (N_{BS})}$ also suggests that an 
RG flow from every point in the critical surface flows into one 
RG flow (RT) in finite $\beta$, 
on which $F(C)^{\rm (N_{BS})}$ is $F(C)^{(\infty)}$, 
as shown in Fig.~\ref{fig:RG},
since the starting point $c_i^{s/t}$ of the infinite block spin 
transformations for $F(C)^{(\infty)}$ must be in $\beta=\infty$ $(g=0)$,

\section{Anisotropic RG-improved action}
\label{sec:ARG}

We search for an action which reproduces the values of Wilson loops 
in the $N_{\rm BS} \to \infty$ limit as much as possible 
within the restricted coupling parameter space 
of the action, Eq.(\ref{eq:gaction}).
For this purpose, 
Iwasaki considered the average relative deviation of Wilson loops,
\begin{eqnarray}
R^{(N_{\rm BS})} = \sqrt{ \sum_{C} \left( \frac{F^{( N_{\rm BS})}(C) 
- F^{(\infty)}(C)} {F^{(\infty)}(C)} \right)^2 w(C)},
\label{eq:r}
\end{eqnarray}
where $\sum_{C}$ is over 4 loop shapes up to length 6 --- 
plaquette (\ref{eq:plaq}), $1\times2$ rectangular loop (\ref{eq:rect}), 
chair (\ref{eq:chair}), and 3-dimensional loop (\ref{eq:3dim}) ---
with a uniform weight $w(C) = 1/4$.
Eq.(\ref{eq:r}) means that, when $R^{(N_{\rm BS})}=0.01$, for example, 
the deviation of small Wilson loops from their values 
in the $N_{\rm BS} \to \infty$ limit is about 1\% 
after $N_{\rm BS}$ block spin transformations.

On anisotropic lattices, we generalize Eq.(\ref{eq:r}) by subdividing each 
loop shape into orientations and adopt a uniform weight for each orientation.
Namely, because we have 3 spatial and 3 temporal plaquette orientations, 
we give $w({\rm spatial\,plaquette}) = w({\rm temporal\,plaquette}) = 1/8$.
For $1\times2$ rectangular loops, we have 6 orientations of spatial loops, 
3 orientations of $W_{k 4}^{(2 \times 1)}$, and 
3 orientations of $W_{k 4}^{(1 \times 2)}$.
Therefore, we give 1/8, 1/16, and 1/16 for their weights.
Similarly, we subdivide 12 chair and 4 3-dimensional loop orientations.

Here, we should emphasize that we are trying to reproduce the values 
of $F^{(\infty)}(C)$ for 10 different Wilson loops by controlling 3 coupling 
parameters for $\xi \neq 1$ (4 Wilson loops by 1 parameter for $\xi=1$) 
at the same time, which is a quite non-trivial trial, 
and the value of $R^{(N_{\rm BS})}$ indicates that $F^{(N_{\rm BS})}(C)$ 
does not change within the accuracy of $R^{(N_{\rm BS})}$ 
under block spin transformations, since $F^{(N_{\rm BS})}(C)$ approaches 
to $F^{(\infty)}(C)$ as $N_{\rm BS}$ increases and the change of 
$F^{(N_{\rm BS})}(C)$ is smaller than the difference. 
Therefore, by measuring the indicator $R^{(N_{\rm BS})}$, 
we can check indirectly how ``slowly'' the coupling parameters flow, 
i.e. how the nearest point which we find in the restricted parameter 
space is close to the real fixed point in the weak coupling limit. 

\begin{figure}[tb]
  \begin{center}
    \leavevmode
    \epsfxsize=8.3cm \epsfbox{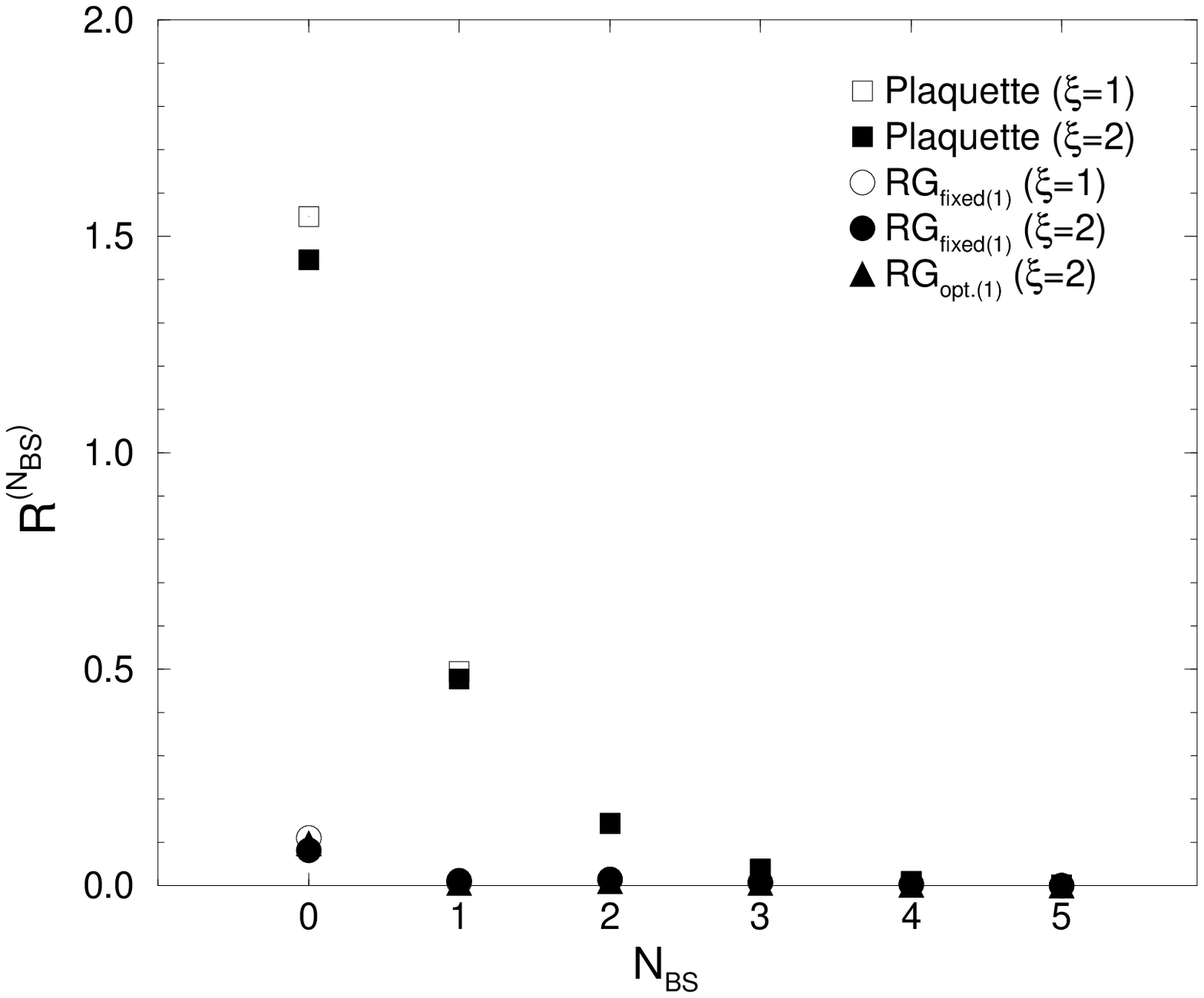}
\hfill
    \epsfxsize=8.3cm \epsfbox{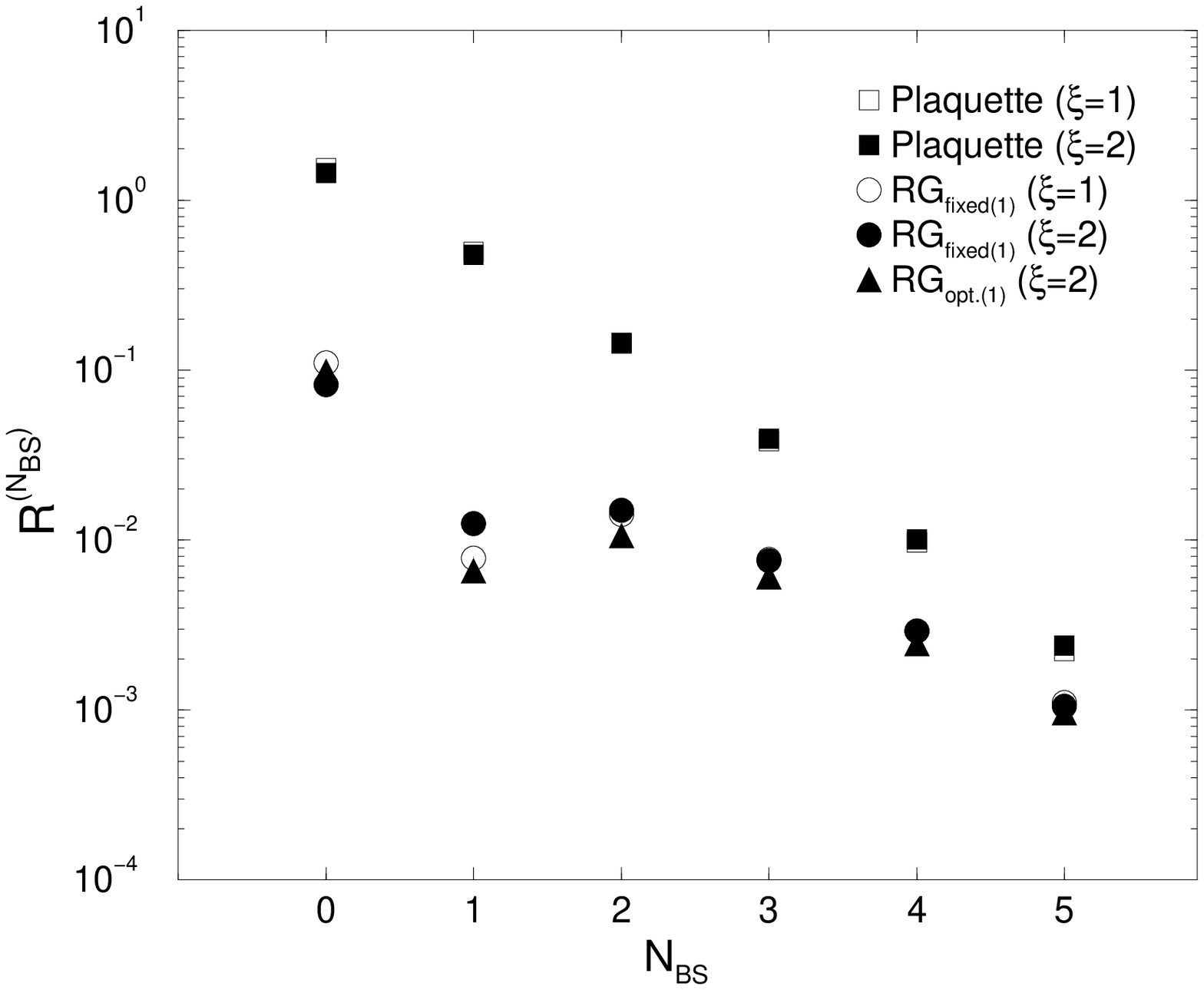}
\caption{$R^{(N_{\rm BS})}$ vs.\ $N_{\rm BS}$ at $\xi=1$ and 2 
for various actions. 
Open symbols are for $\xi=1$ and filled symbols are for $\xi=2$. 
Results from the standard one-plaquette action are shown by squares. 
RG$_{{\rm opt}(n)}$ is the RG-improved action which minimizes 
$R^{(n)}$ on the lattice with the anisotropy $\xi$.
RG$_{{\rm fixed}(n)}$ is an approximate RG-improved action
using the values of $c_i^{s/t}$ for $\xi=1$.}
\label{fig:RBS1}
  \end{center}
\end{figure}

In Fig.~\ref{fig:RBS1}, we show the $N_{\rm BS}$-dependence of 
$R^{(N_{\rm BS})}$ for $\xi=1$ and 2. 
The results from the plaquette action (open and filled squares) show 
exponential decrease with $N_{\rm BS}$. 
(Results from RG-improved actions will be discussed later.)

\begin{figure}[tb]
  \begin{center}
    \leavevmode
    \epsfxsize=8.8cm \epsfbox{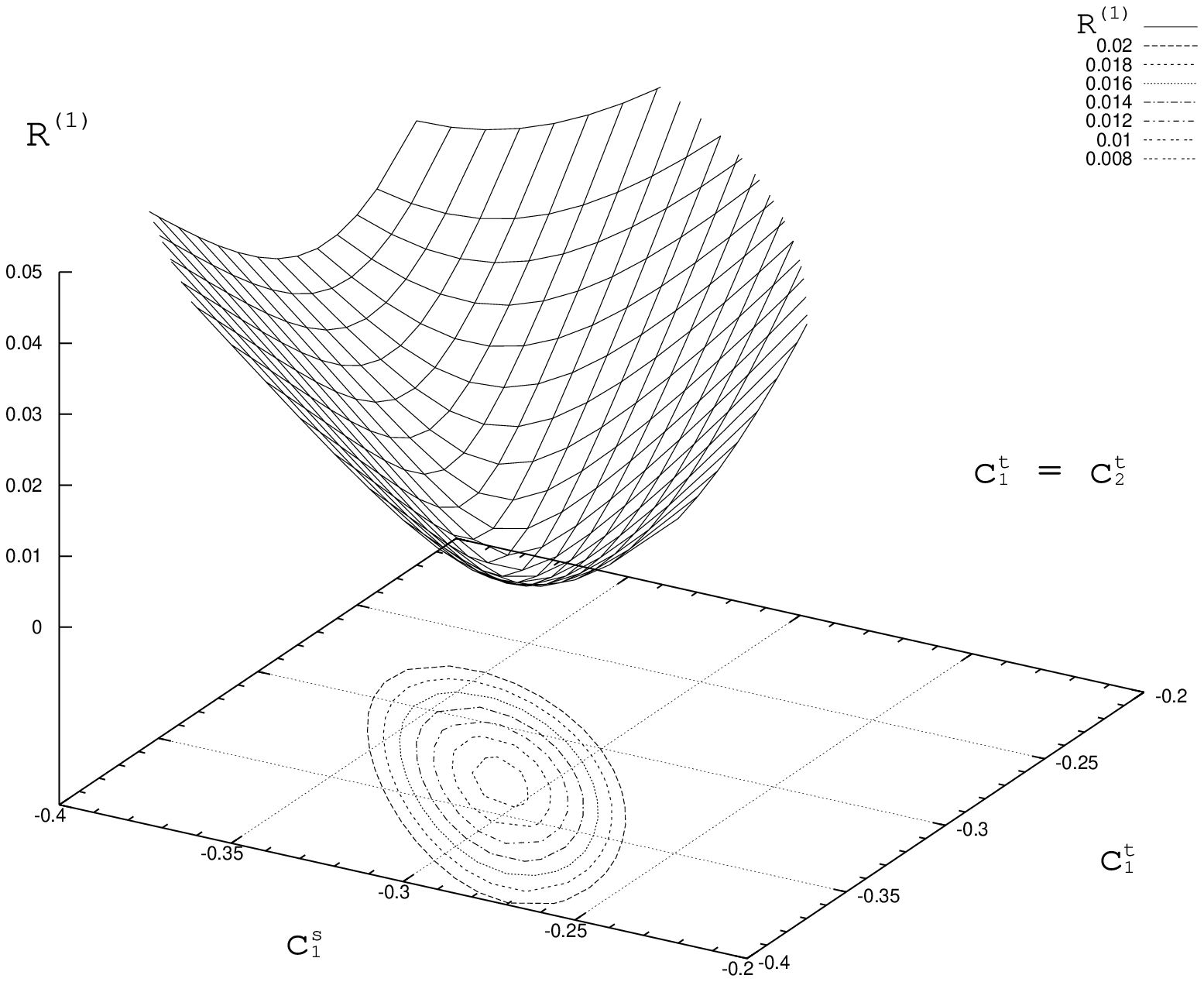}
\hfill
    \epsfxsize=8.8cm \epsfbox{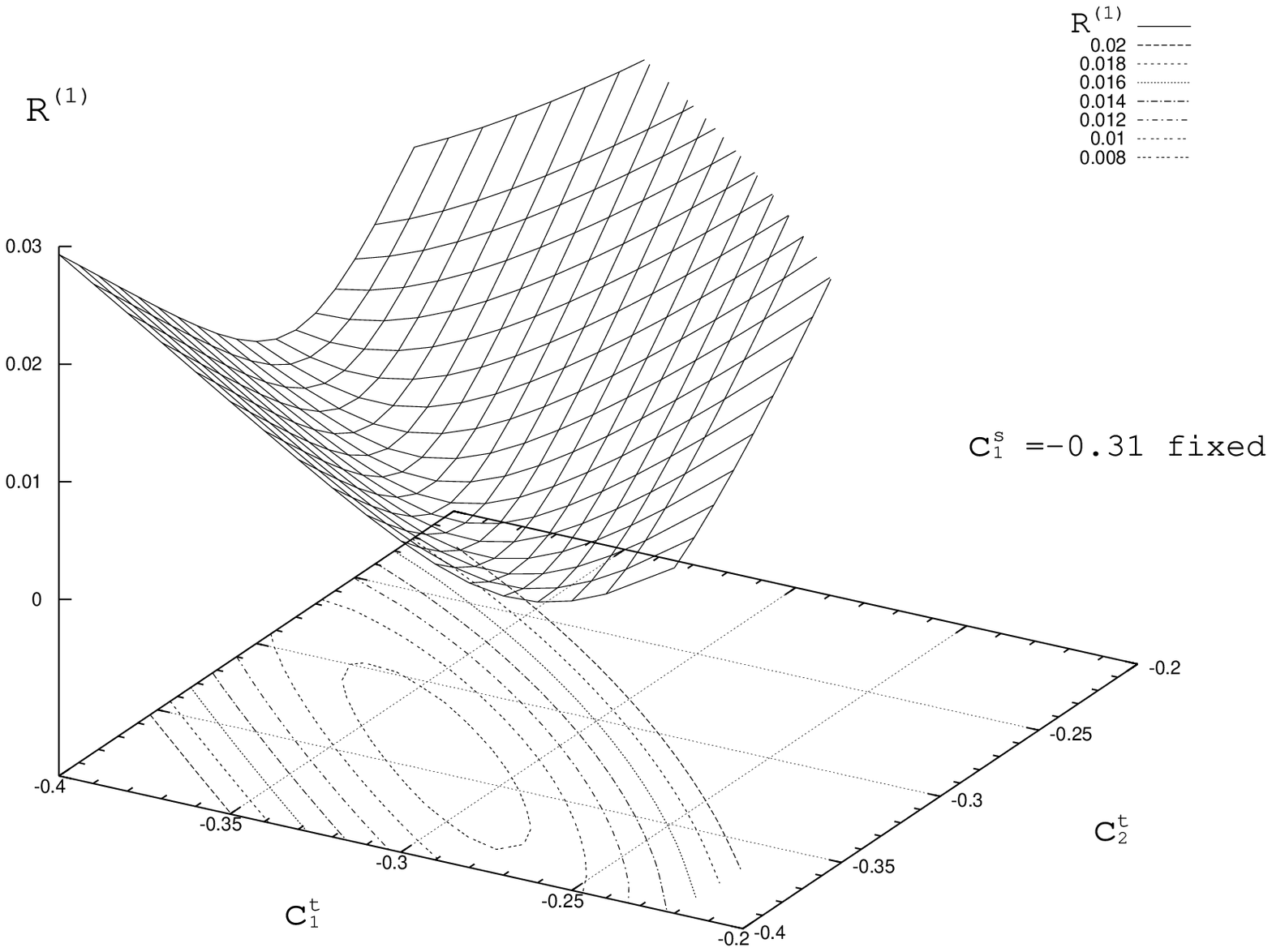}
\caption{$R^{(1)}$ for $\xi=2$; 
(a) in the subspace $c_1^t = c_2^t$ as a function of $(c_1^s, c_1^t = c_2^t)$,
(b) in the subspace $c_1^s = -0.31$ as a function of $(c_1^t, c_2^t)$.
}
\label{fig:3d}
  \end{center}
\end{figure}

We search for the minimum point of $R^{(N_{\rm BS})}$ in the parameter space 
$(c_1^s, c_1^t, c_2^t)$ for each value of $\xi$.
Figure~\ref{fig:3d} shows the behavior of 
$R^{(1)}$ for $\xi=2$ in the subspaces $c_1^t = c_2^t$ and $c_1^s=-0.31$. 
Fig.~\ref{fig:3d}(b) suggests that the region of small $R^{(N_{\rm BS})}$ 
spreads in the direction of constant $c_1^t + c_2^t$, which we confirm 
also for other cases. 

To find the minimum of $R^{(N_{\rm BS})}$, we solve the equations
\begin{eqnarray}
\frac{\partial (R^{(N_{\rm BS})})^2}{\partial c_{i}} = 
\sum_{C} 2 \frac{\partial F^{(N_{\rm BS})}(C)}{\partial c_{i}} 
\frac{F^{(N_{\rm BS})} (C) - F^{(\infty)} (C)}{(F^{(\infty)} (C))^2} w(C) = 0,
\label{eq:min}
\end{eqnarray}
with $c_i = \{ c_1^s, c_1^t$, $c_2^t\}.$
We iteratively solve (\ref{eq:min}) using linear approximations 
\begin{eqnarray}
&& \hspace{-1cm}
\sum_{C} \left( \left. \frac{\partial F^{(N_{\rm BS})}(C)}{\partial c_{i}} 
\right|_{(c^s_{10}, c^t_{10}, c^t_{20})} \right)^2 
\frac{w(C)}{(F^{(\infty)}(C))^2} (c_i -c_{i0}) \nonumber \\
&=& \sum_{C} \left. \frac{\partial F^{(N_{\rm BS})}(C)}{\partial c_{i}} 
\right|_{(c^s_{10}, c^t_{10}, c^t_{20})} \frac{ F^{(\infty)} (C) 
 - \left. F^{(N_{\rm BS})} (C)\right|_{(c^s_{10}, c^t_{10}, c^t_{20})}}
{(F^{(\infty)} (C))^2} w(C),
\end{eqnarray}
around $(c_{10}^s,c_{10}^t,c_{20}^t)$, where 
$\partial F^{(N_{\rm BS})}(C) /\partial c_i$
can be calculated by
\begin{eqnarray}
\frac{\partial D_{\mu \nu}}{\partial c_i} = -D_{\mu \nu} 
\frac{\partial D_{\mu \nu}^{-1}}{\partial c_i} D_{\mu \nu}
\end{eqnarray}
with $D_{\mu \nu}^{-1}$ given by Eq.~(\ref{eq:dinv}).
We solve the equations numerically on a $128^4$ lattice.
We checked that the finite volume effects are sufficiently small 
for the Wilson loops in Eq.(\ref{eq:r}).

\begin{figure}[tb]
  \begin{center}
    \leavevmode
    \epsfxsize=12cm \epsfbox{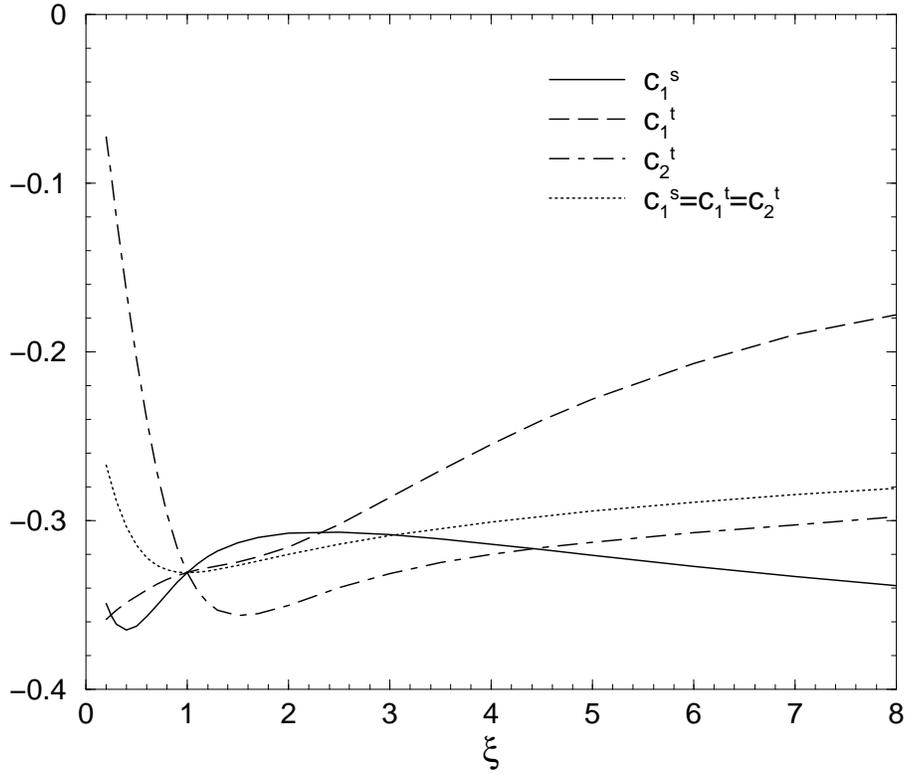}
    \caption{Improvement parameters $(c_1^s, c_1^t, c_2^t)$ for the 
RG-improved action RG$_{{\rm opt}(1)}$ which minimizes $R^{(1)}$ 
at each $\xi$.
The dotted line is the solution which minimizes $R^{(1)}$ when a constraint 
$c_1^s = c_1^t = c_2^t$ is required.}
    \label{fig:bs1}
  \end{center}
\end{figure}

\begin{figure}[tb]
  \begin{center}
    \leavevmode
    \epsfxsize=12cm \epsfbox{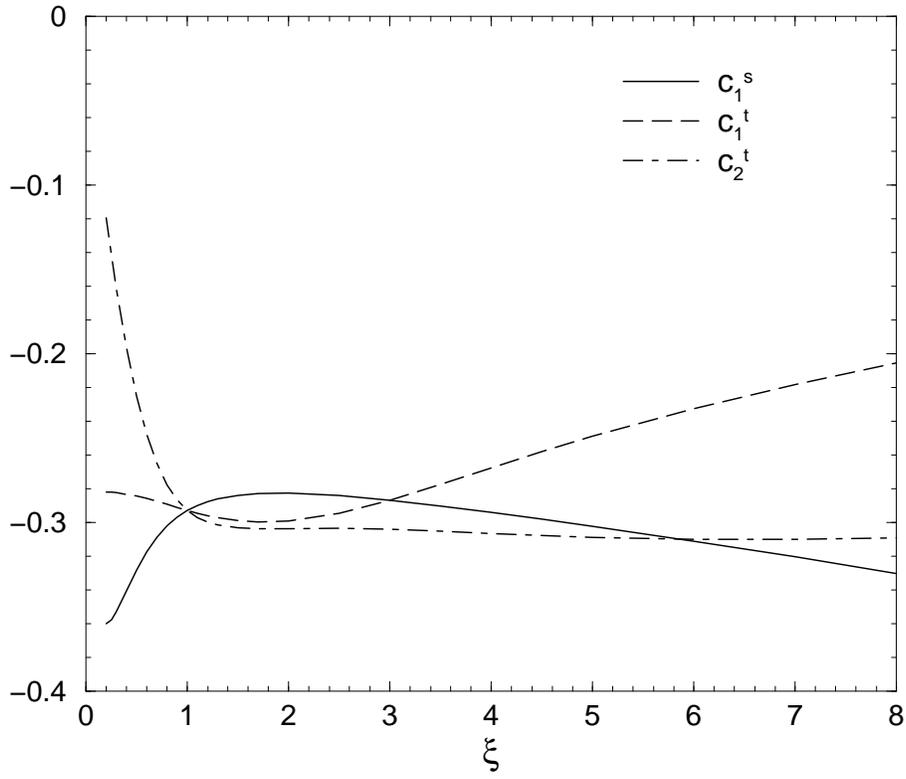}
    \caption{Improvement parameters $(c_1^s, c_1^t, c_2^t)$ for the 
RG-improved action RG$_{{\rm opt}(2)}$ which minimizes $R^{(2)}$ 
at each $\xi$.}
    \label{fig:bs2}
  \end{center}
\end{figure}

\begin{figure}[tb]
  \begin{center}
    \leavevmode
    \epsfxsize=12cm \epsfbox{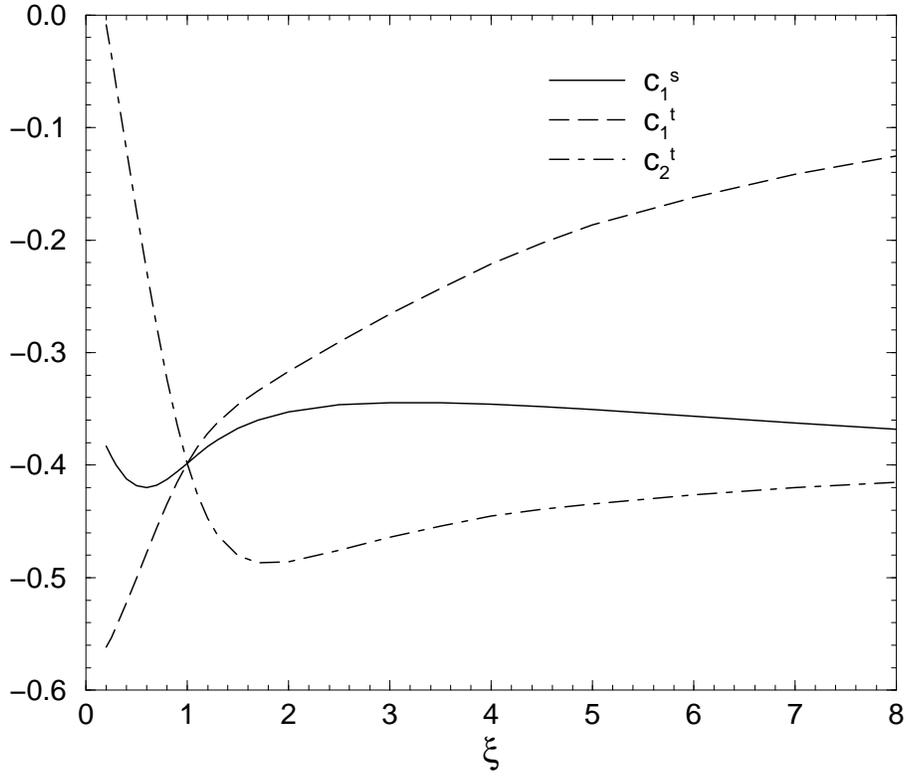}
    \caption{The same as Fig.~\protect{\ref{fig:bs2}}, but for the 
RG$_{{\rm opt}(0)}$ which minimizes $R^{(0)}$.}
    \label{fig:bs0}
  \end{center}
\end{figure}

Results for the improvement parameters which minimize $R^{(N_{\rm BS})}$ 
are summarized in Table~\ref{tab:c1} and 
Figs.~\ref{fig:bs1} and \ref{fig:bs2} for $N_{\rm BS}=1$ and 2. 
We also show the results for $N_{\rm BS}=0$ in Fig.~\ref{fig:bs0}. 
In these figures, 
$c_1^s$, $c_1^t$, and $c_2^t$ are shown by solid, dashed, and dot-dashed 
lines.
In the followings we denote the corresponding action as 
RG$_{{\rm opt}(N_{\rm BS})}$ [the RG-improved action with the 
$\xi$-dependent optimum values of $(c_1^s, c_1^t, c_2^t)$ 
to minimize $R^{(N_{\rm BS})}$].
At $\xi=1$, we reproduce Iwasaki's results \cite{iwasaki}:
$c_1^s = c_i^t = c_2^t = - 0.331$ ($-0.293$) for $N_{\rm BS}=1$ (2).

\begin{figure}[tb]
  \begin{center}
    \leavevmode
    \epsfxsize=12cm \epsfbox{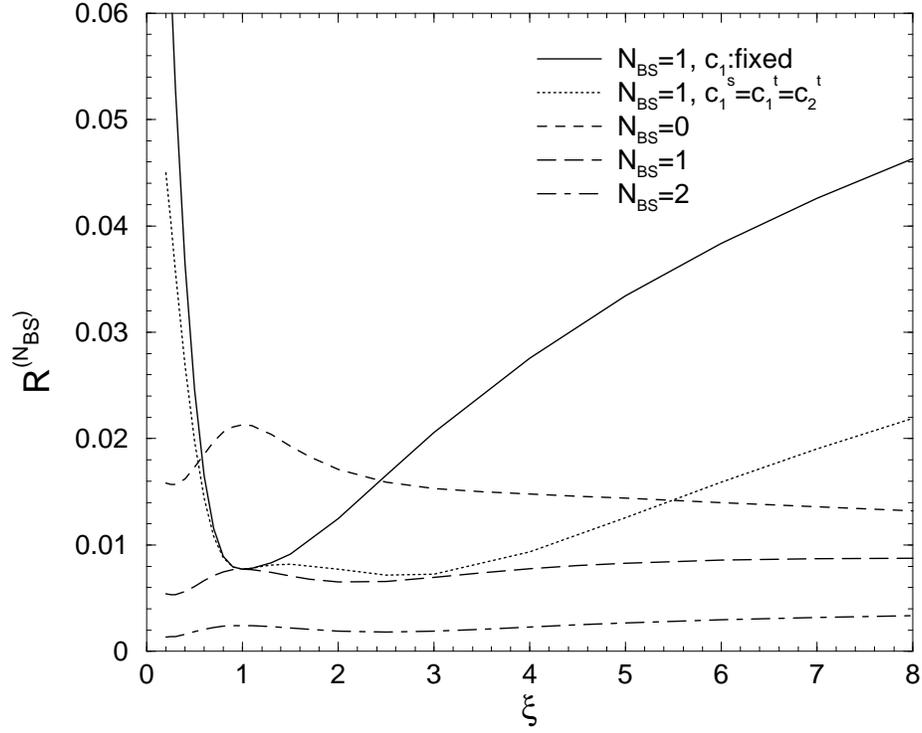}
    \caption{$R^{(N_{\rm BS})}$ as functions of $\xi$ from the RG-improved actions RG$_{{\rm opt}(N_{\rm BS})}$ for $N_{\rm BS}=0$, 1 and 2. 
Also plotted are the results for $R^{(1)}$ determined from 
RG$_{{\rm fixed}(1)}$ (solid line), 
and the minimum $R^{(1)}$ obtained with the constraint 
$c_1^s=c_1^t=c_2^t$ (dotted line). 
}
    \label{fig:r1}
  \end{center}
\end{figure}

Figure \ref{fig:r1} shows the values of $R^{(N_{\rm BS})}$ from 
RG$_{{\rm opt}(N_{\rm BS})}$ for $N_{\rm BS}=0,$ 1 and 2 
(dashed, dot dashed, dot dot dashed lines, respectively).
We find that the values of $R^{(N_{\rm BS})}$ remain small in 
a wide range of $\xi$, 
indicating that a similar quality of improvement is achieved 
by the program of RG-improvement even at $\xi \neq 1$. 



Here, it is worth noting that reducing the number of independent 
coupling parameters has a practical benefit in numerical simulations. 
In particular, because a non-trivial $\xi$-dependence in coupling parameters 
makes the calculation of thermodynamic quantities complicated,
it is attractive to adopt $\xi$-independent improvement parameters. 

Therefore, we study $R^{(N_{\rm BS})}$ at $\xi \neq 1$ with 
the improved parameters fixed to the optimum value at $\xi = 1$,  
$c_1^s = c_1^t = c_2^t (= -0.331 \; {\rm for} \; N_{\rm BS}=1)$. 
We denote this action as RG$_{{\rm fixed}(N_{\rm BS})}$ 
[an RG-improved action
with $c_1^s$, $c_1^t$, and $c_2^t$ fixed to the Iwasaki's value 
minimizing $R^{(N_{\rm BS})}$ on the $\xi=1$ lattice].
The result of $R^{(1)}$  
for $N_{\rm BS}=1$ is plotted by the solid line in Fig.~\ref{fig:r1}. 
We also study the case $c_1^s = c_1^t = c_2^t \equiv c_1(\xi)$ 
where $c_1$ is varied to minimize $R^{(N_{\rm BS})}$ at each $\xi$. 
The results for the minimum value of $R^{(1)}$ and 
the corresponding optimum value of the parameter $c_1$ 
are shown by dotted lines in Figs.~\ref{fig:r1} and \ref{fig:bs1}, 
respectively.
For both cases, $R^{(1)}$ becomes larger as $\xi$ deviates from 1.
It means that one cannot keep the same quality of improvement 
in the whole range of $\xi$ 
with the RG$_{{\rm fixed}(N_{\rm BS})}$ action 
nor the action with the constraint $c_1^s = c_1^t = c_2^t = c_1(\xi)$. 

In most simulations, however, we are interested in the cases of 
$\xi\approx1$--4, where the values of $R^{(1)}$ remain $O(10^{-2})$.
In the determination of an RG-improved action, the difference between 
Fig.~\ref{fig:bs1} for $N_{\rm BS}=1$ and Fig.~\ref{fig:bs2} for 
$N_{\rm BS}=2$ is a matter of taste: 
Both actions are equally qualified with $R^{(N_{\rm BS})}\leq O(10^{-2})$
and the difference in the values of improvement parameters should be 
regarded as a freedom in the choice. 
In this respect, we find that the variations of improvement parameters 
as functions of $\xi$ are small for $\xi\approx1$--4. 

\begin{figure}[tb]
  \begin{center}
    \leavevmode
    \epsfxsize=12cm \epsfbox{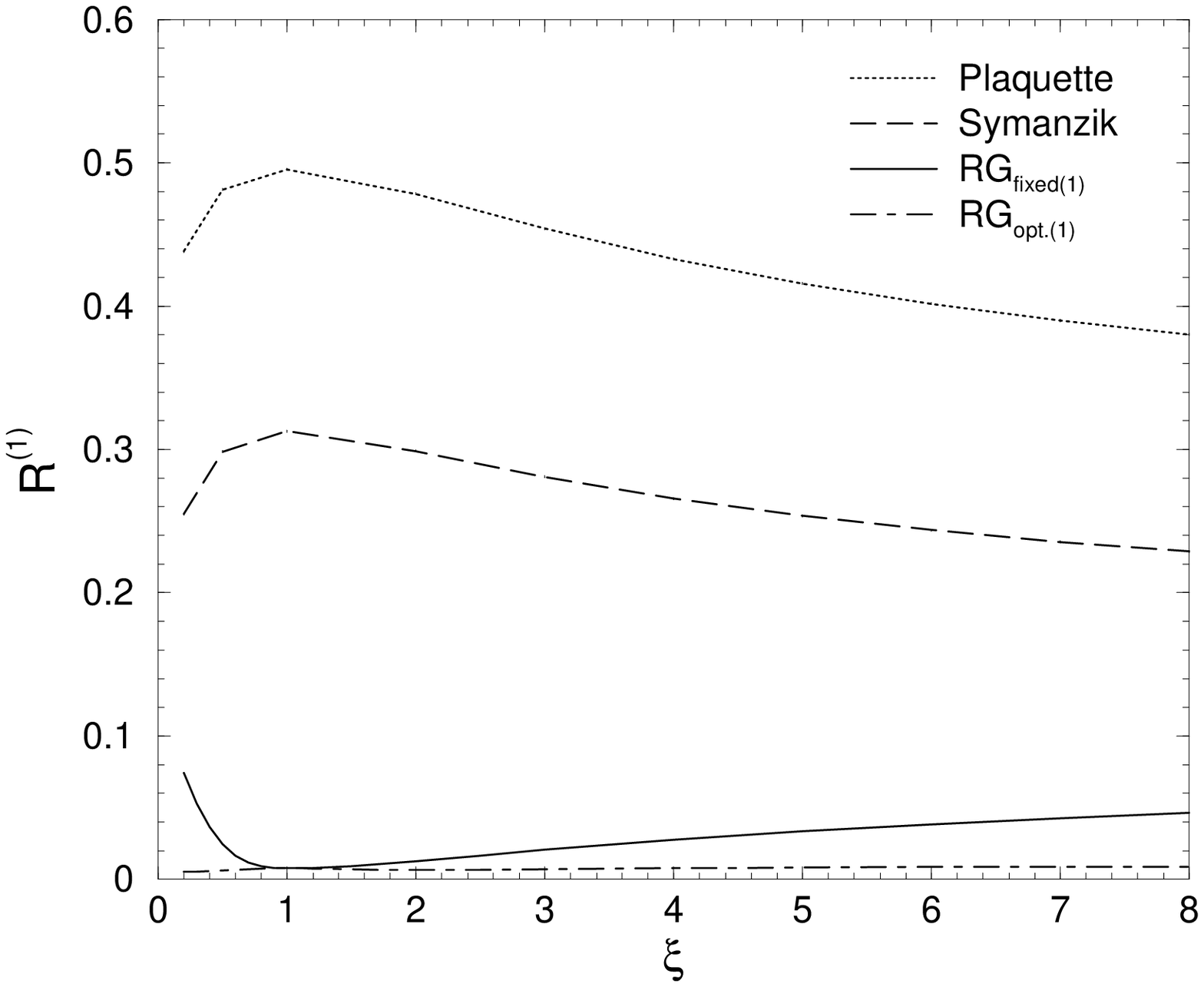}
    \caption{$R^{(1)}$ for various actions as functions of $\xi$. 
}
    \label{fig:r2}
  \end{center}
\end{figure}

In Fig.~\ref{fig:r2}, we show $R^{(1)}$ for various actions 
including the standard plaquette action and the Symanzik improved actions. 
For RG-improved actions, results are shown for 
RG$_{{\rm opt}(1)}$ and RG$_{{\rm fixed}(1)}$.
Similar results are obtained for other values of $N_{\rm BS}$, too.
We find that, although a stable improvement is achieved with the 
RG$_{{\rm opt}(1)}$ action for a wide range of $\xi$, 
when we restrict ourselves to the range $\xi\approx1$--4, 
all RG-improved actions lead to quite small values of 
$R^{(1)}$ \simlt $O(10^{-2})$, 
i.e. the average deviation of small Wilson loops from the 
$N_{\rm BS} = \infty$ limit is less than about 1\% after one blocking. 
On the other hand, for the standard plaquette and Symanzik actions, 
typical values of $R^{(1)}$ are 0.4--0.5 and 0.25--0.3, respectively. 
We conclude that the RG$_{{\rm fixed}(N_{\rm BS})}$ action, 
in which the improvement parameters are fixed to Iwasaki's 
value for $\xi=1$, improves the theory well at $\xi\approx1$--4. 


\section{Conclusions}
\label{sec:conc}

We studied RG-improved actions for the SU(3) gauge theory on 
anisotropic lattices, following Iwasaki's program of improvement. 
We determined the improvement parameters as functions of the 
anisotropy $\xi$ for the action with plaquette and $1\times2$ rectangular terms. 
We found that the program of improvement works well even on anisotropic 
lattices without losing the quality of improvement if we adjust three 
improvement parameters $(c_1^s, c_1^t, c_2^t)$ as functions of $\xi$. 

Moreover, we discussed a practical choice of improved action 
for numerical simulations on anisotropic lattices. 
From a calculation of an indicator which estimates the distance to 
the renormalized trajectory near the fixed point,
we found that keeping the improvement parameters to the values at
$\xi=1$ leads to the distance comparable to the minimum distance 
at the optimum $(c_1^s, c_1^t, c_2^t)$, 
for the range of the anisotropy $\xi \approx 1$--4.
This means that, for the range $\xi\approx 1$--4 
where anisotropic lattices are expected to be efficient in calculating
thermodynamic quantities \cite{name01},
the choice of Iwasaki's value for improvement parameters is
acceptable also for $\xi \neq 1$, 
as adopted in a previous work \cite{sakai}.

As the next step, it is necessary to confirm whether good properties 
of the RG-improved action at $\xi=1$ maintain also at $\xi \neq 1$ 
in practical simulations, but we showed that the Iwasaki's program of 
an RG-improved action can be generalized for $\xi \neq 1$.

\section*{Acknowledgments}
We thank the members of the CP-PACS Collaboration and S. Hands 
for helpful comments and discussions. 
This work is supported in part by Grants-in-Aid of the Ministry of 
Education, Culture, Sports, Science and Technology, Japan 
(No.~13640260).
SE is supported by PPARC grant PPA/G/S/1999/00026. 
TU is supported by the public research organizations at 
Center for Computational Physics (CCP), University of Tsukuba.

\section*{Appendix A: Derivation of the Wilson loop after 
block spin transformation}
\label{sec:appenA}

We derive Eqs.(\ref{eq:fbij}), (\ref{eq:fbch}) and (\ref{eq:fb3d})
following the appendix B of Ref.~\cite{iwasaki}.
Let us introduce the Fourier transformation
\begin{eqnarray}
A_\mu^{(N_{\rm BS})} (x)=\int_k {\rm e}^{i2^{N_{\rm BS}} (kx +k_{\mu}/2)} 
\tilde{A}^{(N_{\rm BS})}_{\mu}(k),
\label{eq:abs}
\end{eqnarray}
where the lattice spacing for $A^{(N_{\rm BS})}_{\mu}(x)$ is 
$2^{N_{\rm BS}} a$.
From Eqs.~(\ref{eq:four}), (\ref{eq:block}) and (\ref{eq:abs}), we obtain 
\begin{eqnarray}
\tilde{A}_\mu^{(N_{\rm BS})} (k) = {\rm e}^{-i(2^{N_{\rm BS}-1} -1/2)
k_{\mu}} \tilde{H}^{(N_{\rm BS})} (k) \tilde{A}_{\mu}^{(0)} (k),
\end{eqnarray}
where
\begin{eqnarray}
\tilde{H}^{(N_{\rm BS})} (k) = \prod_{M=0}^{N_{\rm BS}-1} 
\frac{1}{8} \prod_{\nu=1}^{4} ({\rm e}^{i2^M k_{\nu}}+1).
\end{eqnarray}
We define the free propagator $D_{\mu \nu}^{(N_{\rm BS})}$ for 
the field $\tilde{A}^{(N_{\rm BS})}$ by
\begin{eqnarray}
\langle \tilde{A}_{\mu}^{a (N_{\rm BS})} (k) 
\tilde{A}_{\nu}^{b (N_{\rm BS})} (k') \rangle 
= \delta_{a,b} \delta (k+k') D^{(N_{\rm BS})}_{\mu \nu} (k).
\end{eqnarray}
We obtain
\begin{eqnarray}
D_{\mu \nu}^{(N_{\rm BS})} (k) = {\rm e}^{-i(2^{N_{\rm BS}-1} -1/2)k_{\mu}} 
{\rm e}^{i(2^{N_{\rm BS}-1} -1/2)k_{\nu}} H^{(N_{\rm BS})} (k) 
D_{\mu \nu} (k),
\end{eqnarray}
where $H^{(N_{\rm BS})} (k)$ is given by Eq.~(\ref{eq:hbs}).

The expectation value of Wilson loops that can be written as
\begin{eqnarray}
W(C) = \sum_{\mu, \nu} c_{\mu \nu} (k) D_{\mu \nu} (k)
\end{eqnarray}
for the original lattice is obtained for the $N_{\rm BS}$-th blocked lattice by
\begin{eqnarray}
W^{(N_{\rm BS})} (C) = \sum_{\mu, \nu} c_{\mu \nu} (2^{N_{\rm BS}} k) 
D^{(N_{\rm BS})}_{\mu \nu} (k)
\end{eqnarray}
To derive Eq.~(\ref{eq:hbs}), we used the fact that 
$D_{\mu \nu}(k)$ is odd in $k_{\mu}$ and $k_{\nu}$ when $\mu \neq \nu$.


\newpage
\begin{table}[tb]
\caption{Blocked Wilson loops and those $N^{\rm BS} \to \infty$ limit. 
$F_s$ and $F_t$ are for spatial and space-time Wilson loops, respectively.
$N_s^3 \times N_t = 128^3 \times (128 \xi)$
}
\label{tab:wl}
\begin{tabular}{ccccc}
$\xi=1$ &&& \\
\hline
$N^{\rm BS}$ & $F(1 \times 1)$ & $F(1 \times 2)$ & 
& $F(2 \times 2)$ \\
\hline
0 & 0.500000 & 0.862251 & & 1.369312 \\
1 & 0.288104 & 0.517653 & & 0.879783 \\
2 & 0.216234 & 0.403513 & & 0.720860 \\
3 & 0.194450 & 0.369800 & & 0.674938 \\
4 & 0.188403 & 0.360256 & & 0.660681 \\
$\infty$ & 0.186476 & 0.357678 & & 0.658761 \\
\hline
\hline
$\xi=2$ &&& \\
\hline
$N^{\rm BS}$ & $F_s(1 \times 1)$ & $F_s(1 \times 2)$ 
& $F_s(2 \times 1)$ & $F_s(2 \times 2)$ \\
\hline
0 & 0.673095 & 1.128029 & 1.128029 & 1.728563 \\
1 & 0.402015 & 0.701155 & 0.701155 & 1.145925 \\
2 & 0.307103 & 0.555452 & 0.555452 & 0.952321 \\
3 & 0.277085 & 0.510586 & 0.510586 & 0.894124 \\
4 & 0.268545 & 0.497689 & 0.497689 & 0.876158 \\
$\infty$ & 0.265709 & 0.493899 & 0.493899 & 0.872921 \\
\hline
$N^{\rm BS}$ & $F_t(1 \times 2)$ & $F_t(1 \times 4)$ 
& $F_t(2 \times 2)$ & $F_t(2 \times 4)$ \\
\hline
0 & 0.556383 & 0.920967 & 0.995852 & 1.510346 \\
1 & 0.348216 & 0.572014 & 0.655826 & 1.005403 \\
2 & 0.274425 & 0.454023 & 0.537939 & 0.837727 \\
3 & 0.251345 & 0.418526 & 0.501772 & 0.788069 \\
4 & 0.244845 & 0.408394 & 0.491385 & 0.772639 \\
$\infty$ & 0.242740 & 0.405580 & 0.488471 & 0.770227 \\
\end{tabular}
\end{table}

\begin{table}[tb]
\caption{The improvement parameters $(c_1^s, c_1^t, c_2^t)$ 
and $R^{(N_{\rm BS})}$ of the RG-improved action RG$_{{\rm opt}(N_{\rm BS})}$
for $N_{\rm BS}=1$ and 2 at various $\xi$.
}
\label{tab:c1}
\begin{tabular}{cccccc}
$N_{\rm BS}$ & $\xi$ & $c_1^s$ & $c_1^t$ & $c_2^t$ & $R^{(N_{\rm BS})}$ \\
\hline
1 & 0.20 & $-$0.349 & $-$0.359 & $-$0.073 & 5.40 $\times 10^{-3}$ \\
1 & 0.25 & $-$0.357 & $-$0.356 & $-$0.097 & 5.33 $\times 10^{-3}$ \\
1 & 0.50 & $-$0.363 & $-$0.345 & $-$0.204 & 6.10 $\times 10^{-3}$ \\
1 & 0.70 & $-$0.350 & $-$0.338 & $-$0.271 & 7.14 $\times 10^{-3}$ \\
1 & 0.90 & $-$0.337 & $-$0.333 & $-$0.316 & 7.67 $\times 10^{-3}$ \\
1 & 1.00 & $-$0.331 & $-$0.331 & $-$0.331 & 7.73 $\times 10^{-3}$ \\
1 & 1.10 & $-$0.326 & $-$0.329 & $-$0.341 & 7.68 $\times 10^{-3}$ \\
1 & 1.50 & $-$0.313 & $-$0.324 & $-$0.356 & 7.07 $\times 10^{-3}$ \\
1 & 2.00 & $-$0.307 & $-$0.316 & $-$0.350 & 6.52 $\times 10^{-3}$ \\
1 & 2.50 & $-$0.307 & $-$0.302 & $-$0.340 & 6.58 $\times 10^{-3}$ \\
1 & 3.00 & $-$0.308 & $-$0.286 & $-$0.331 & 6.96 $\times 10^{-3}$ \\
1 & 3.50 & $-$0.311 & $-$0.270 & $-$0.325 & 7.39 $\times 10^{-3}$ \\
1 & 4.00 & $-$0.314 & $-$0.255 & $-$0.320 & 7.77 $\times 10^{-3}$ \\
1 & 4.50 & $-$0.317 & $-$0.241 & $-$0.316 & 8.07 $\times 10^{-3}$ \\
1 & 5.00 & $-$0.320 & $-$0.229 & $-$0.312 & 8.30 $\times 10^{-3}$ \\
1 & 6.00 & $-$0.327 & $-$0.208 & $-$0.306 & 8.59 $\times 10^{-3}$ \\
1 & 7.00 & $-$0.333 & $-$0.192 & $-$0.302 & 8.73 $\times 10^{-3}$ \\
1 & 8.00 & $-$0.339 & $-$0.178 & $-$0.298 & 8.77 $\times 10^{-3}$ \\
\hline
2 & 0.20 & $-$0.360 & $-$0.282 & $-$0.122 & 1.34 $\times 10^{-3}$ \\
2 & 0.25 & $-$0.358 & $-$0.282 & $-$0.142 & 1.37 $\times 10^{-3}$ \\
2 & 0.50 & $-$0.328 & $-$0.284 & $-$0.225 & 1.87 $\times 10^{-3}$ \\
2 & 0.70 & $-$0.309 & $-$0.287 & $-$0.265 & 2.25 $\times 10^{-3}$ \\
2 & 0.90 & $-$0.297 & $-$0.291 & $-$0.287 & 2.42 $\times 10^{-3}$ \\
2 & 1.00 & $-$0.293 & $-$0.293 & $-$0.293 & 2.43 $\times 10^{-3}$ \\
2 & 1.10 & $-$0.290 & $-$0.294 & $-$0.297 & 2.42 $\times 10^{-3}$ \\
2 & 1.50 & $-$0.284 & $-$0.299 & $-$0.303 & 2.20 $\times 10^{-3}$ \\
2 & 2.00 & $-$0.283 & $-$0.299 & $-$0.304 & 1.91 $\times 10^{-3}$ \\
2 & 2.50 & $-$0.284 & $-$0.295 & $-$0.303 & 1.82 $\times 10^{-3}$ \\
2 & 3.00 & $-$0.287 & $-$0.287 & $-$0.304 & 1.91 $\times 10^{-3}$ \\
2 & 3.50 & $-$0.290 & $-$0.277 & $-$0.305 & 2.09 $\times 10^{-3}$ \\
2 & 4.00 & $-$0.294 & $-$0.268 & $-$0.306 & 2.30 $\times 10^{-3}$ \\
2 & 4.50 & $-$0.298 & $-$0.258 & $-$0.308 & 2.50 $\times 10^{-3}$ \\
2 & 5.00 & $-$0.302 & $-$0.249 & $-$0.309 & 2.68 $\times 10^{-3}$ \\
2 & 6.00 & $-$0.311 & $-$0.232 & $-$0.310 & 2.98 $\times 10^{-3}$ \\
2 & 7.00 & $-$0.320 & $-$0.218 & $-$0.310 & 3.20 $\times 10^{-3}$ \\
2 & 8.00 & $-$0.330 & $-$0.205 & $-$0.309 & 3.36 $\times 10^{-3}$ \\
\end{tabular}
\end{table}

\newpage

\begin{thebibliography}{99}


\bibitem{cppacsFull} 
A. Ali Khan {\it et al.} (CP-PACS Collaboration), 
Phys.\ Rev.\ D65 (2002) 054505;
A. Ali Khan {\it et al.} (CP-PACS Collaboration), 
Phys.\ Rev.\ Lett.\ 85 (2000) 4674;

\bibitem{comparative}
S.\ Aoki {\it et al.} (CP-PACS Collaboration), 
Phys.\ Rev.\ D60 (1999) 114508.

\bibitem{tsukuba97}
Y. Iwasaki, K. Kanaya, S. Kaya, and T. Yoshi\'{e}, 
Phys. Rev. Lett. 78 (1997) 179. 

\bibitem{cppacs00}
A. Ali Khan {\it et al.} (CP-PACS Collaboration), 
Phys.\ Rev.\ D63 (2000) 034502.

\bibitem{milc} 
C.\ Bernard {\it et al.}, Phys.\ Rev.\ D58 (1998) 014503;
Nucl.\ Phys.\ B (Proc.\ Suppl.) 106 (2002) 412.

\bibitem{p4action} 
U.M.\ Heller, F.\ Karsch, and B.\ Sturm, 
Phys.\ Rev.\ D60 (1999) 114502; 
F.\ Karsch, E.\ Laermann, and A.\ Peikert, 
Phys.\ Lett.\ B478 (2000) 447; 
Nucl.\ Phys.\ B605 (2001) 579.

\bibitem{milc97} 
C.\ Bernard {\it et al.}, Phys.\ Rev.\ D55 (1997) 6861.

\bibitem{cppacs01}
A. Ali Khan {\it et al.} (CP-PACS Collaboration), 
Phys.\ Rev.\ D64 (2001) 074510.

\bibitem{ejiri01}
S.\ Ejiri, Nucl.\ Phys.\ B (Proc.\ Suppl.) 94 (2001) 19.

\bibitem{name01} 
Y. Namekawa {\it et al.} (CP-PACS Collaboration), 
Phys.\ Rev.\ D64 (2001) 074507.

\bibitem{sakai}
S.\ Sakai, A.\ Nakamura and T.\ Saito,
Nucl.\ Phys.\ {\bf A638} (1998) 535;
S. Sakai, T. Saito, and A. Nakamura, 
Nucl.\ Phys.\ B584 (2000) 528.

\bibitem{taro}
Ph.\ de Forcrand {\it et al.} (QCD-TARO Collaboration), 
Phys.\ Rev.\ {\bf D63} (2001) 054501.

\bibitem{umeda}
T.\ Umeda, R.\ Katayama, O.\ Miyamura and H.\ Matsufuru,
Nucl.\ Phys.\ {\bf B} (Proc.\ Suppl.) {\bf 94} (2001) 435; 
hep-lat/0011085.

\bibitem{klassen}
T.R.\ Klassen, Nucl.\ Phys.\ B (Proc.\ Suppl.) 73 (1999) 918;
P.\ Chen, Phys.\ Rev.\ D64 (2001) 034509.

\bibitem{bali}
G.S.\ Bali and P.\ Boyle, Phys.\ Rev.\ D59 (1999) 114504;
G.S.\ Bali, Phys.\ Rev.\ D62 (2000) 114503.

\bibitem{okamoto}
M.\ Okamoto {\it et al.} (CP-PACS Collaboration),
hep-lat/0112020 [Phys.\ Rev.\ D in press].

\bibitem{manke}
T.\ Manke {\it et al.} (CP-PACS Collaboration),
Phys.\ Rev.\ Lett.\ 82 (1999) 4396.

\bibitem{morningstar}
C.J.\ Morningstar and M.\ Peardon,
Phys.\ Rev.\ D56 (1997) 4043;
Phys.\ Rev.\ D60 (1999) 034509.

\bibitem{liu}
C.\ Liu, J.\ Zhang, Y.\ Chen, J.P.\ Ma, Nucl.\ Phys.\ B624 (2002) 360.

\bibitem{iwasaki} 
Y.\ Iwasaki, Nucl.\ Phys.\ B258 (1985) 141; 
Univ.\ of Tsukuba report UTHEP-118 (1983) unpublished.

\bibitem{karsch81} 
F.\ Karsch, Nucl. Phys. B205 (1982) 285.

\bibitem{burgers88}
G.\ Burgers, F.\ Karsch, A.\ Nakamura and I.O.\ Stamaescu,
Nucl.\ Phys.\ B304 (1988) 587.

\bibitem{klassen98}
T.R.\ Klassen, Nucl. Phys. B533 (1998) 557.

\bibitem{ejiri98}
S.\ Ejiri, Y.\ Iwasaki and K.\ Kanaya, Phys. Rev. D 58 (1998) 094505.

\bibitem{engels00}
J.\ Engels, F.\ Karsch and T.\ Scheideler, Nucl. Phys. B564 (2000) 303.

\bibitem{alford98}
M.\ Alford, T.R.\ Klassen and G.P.\ Lepage, Phys.\ Rev.\ D58 (1998) 034503.

\bibitem{morningstar97}
C.\ Morningstar, Nucl.\ Phys.\ B (Proc.\ Suppl.) 53 (1997) 914.

\bibitem{alford01}
M.\ Alford, I.T.\ Drummond, R.R.\ Horgan, H.\ Shanahan and M.\ Peardon,
Phys.\ Rev.\ D63 (2001) 074501.

\bibitem{hasenfratz94}
P.\ Harsenfratz and F.\ Niedermayer, Nucl. Phys. B414 (1994) 785.

\bibitem{taro98} 
P.\ de Forcrand {\it et al.} (QCD-TARO Collaboration), 
Nucl.\ Phys.\ B577 (2000) 263.

\bibitem{ruefenacht}
P.\ R\"uefenacht and U.\ Wenger, Nucl.\ Phys.\ B616 (2001) 163.


\bibitem{RGpot}
Y.\ Iwasaki, K.\ Kanaya, T.\ Kaneko and T.\ Yoshi\'e,
Phys. Rev. D 56 (1997) 151.

\bibitem{cppacsDW}
A. Ali Khan {\it et al.} (CP-PACS Collaboration), 
Phys.\ Rev.\ D63 (2001) 114504;
Phys.\ Rev.\ D64 (2001) 114506;
J.-I. Noaki {\it et al.} (CP-PACS Collaboration), 
hep-lat/0108013.

\bibitem{morningstar96}
C.\ Morningstar and M.\ Peardon,
Nucl.\ Phys.\ B (Proc.\ Suppl.) 47 (1996) 258.

\bibitem{weisz83} 
P.\ Weisz, Nucl. B212 (1983) 1.

\end{thebibliography}
\end{document}